\documentclass[journal]{IEEEtran}

\usepackage{url}
\usepackage{amsmath, amstext, amssymb}
\usepackage{graphicx}
\usepackage{psfrag}
\usepackage{multirow}
\usepackage{hhline}
\usepackage{eurosym}
\usepackage{color}

\begin{document}
%
\title{A Fully Controllable Power System | Concept for FACTS and HVDC Placement}

\author{Spyros~Chatzivasileiadis,~\IEEEmembership{Member,~IEEE,}
        and~G\"{o}ran~Andersson,~\IEEEmembership{Fellow,~IEEE}

\thanks{S. Chatzivasileiadis is with Lawrence Berkeley National Laboratory, e-mail:schatzivasileiadis@lbl.gov. G. Andersson is with the EEH-Power Systems Laboratory, Department of Electrical and Computer Engineering, ETH Zurich, Switzerland, e-mail: andersson@eeh.ee.ethz.ch.}}

%



\maketitle

\begin{abstract}
This paper puts forward the vision of fully decoupling market operations from security considerations through controllable power flows. In ``A Fully Controllable Power System'', power system security is no longer dependent on the location of the power injection points. In the ideal case, this leads to the elimination of redispatching costs, which amount to several million dollars per year in large systems. This paper determines the upper and lower bounds for the number of controllable lines and number of controllers to achieve this decoupling in any system. It further introduces the notion of the controllability vector $CV$, which expresses the effect of any controller on the AC line flows. Based on two alternative definitions for controllability, two controller placement algorithms to maximize controllability are presented and their results are compared.  
\end{abstract}


%
\IEEEpeerreviewmaketitle

\section{Introduction}


``A Fully Controllable Power System'' aims to arrive at a system where the market operations are fully decoupled from the security considerations. In such a system, the ``Cost of Security'' will be zero, as the system controllability will guarantee the security of the system without need for any costly preventive control actions, such as the generation redispatching.

The first objective in power system operation is to satisfy the power demand. As long as a Standard OPF results in a feasible solution, the system has both the necessary generation and transmission capacity to supply all loads. Normally, however, most power systems are dispatched in a way to be N-1 secure. For that purpose, a Security-Constrained OPF (SCOPF) is executed. Due to the uncontrollable nature of the AC line flows, the points where the power is injected play a crucial role on the resulting power flows and on possible overloadings. As long as the Security-Constrained OPF results in a feasible solution, the system has additional generation capacity, and, most importantly, sufficient transmission capacity, so that even if an element is lost, no overloadings will occur. By definition, the SCOPF will result in an equal or more expensive dispatch than the Standard OPF. In case we were able to control all line flows, however, the injection points would play a substantially less significant role. 

The ``Cost of Security'' reflects the additional costs incurred to the system, so that N-1 security is ensured \cite{SCH_SEPOPE_2012}. It is calculated as the difference between the system costs resulting from a Security-Constrained OPF minus the system costs resulting from a standard OPF, as shown in \eqref{ch3_equ:CoS_def}.
\begin{equation}
CoS=C_{SCOPF}-C_{OPF}
\label{ch3_equ:CoS_def}
\end{equation}
Several approaches exist in the literature, which aim to take advantage of power system controllability to provide both preventive and corrective control and minimize the redispatching costs. Ref.~\cite{Monticelli_SCOPF_corrective_first} presented the first formulation incorporating corrective control actions from generation rescheduling and switching actions. In \cite{Hedman_TransSwitching} the benefits of transmission switching for optimizing the power system economic operation are demonstrated. In \cite{Vittal_OPF_FACTS} optimization algorithms for corrective control with FACTS are presented, while, to our knowledge, \cite{SCH_SEPOPE_2012} presented the first SCOPF formulation with corrective control of HVDC lines. 

The goal of the concept ``A fully Controllable Power System'' is to decouple the dependence of the line flows on the power injection points. 
Market operations, e.g. the solution of a standard OPF, should determine the most economical dispatch. Through fully controllable power flows, system operators will guarantee system security by appropriately rerouting the power before and/or after a contingency occurs. Thus, as long as the transmission capacity is sufficient, there will be no need for (preventive) generator redispatching actions and the dispatch determined through the market is accepted as is. By incuring no additional redispatching costs this results to a zero ``Cost of Security'' in the ideal case.

In the rest of this paper we will deal with two main problems. First, the minimum number of controllers that are necessary in order to make the system fully controllable will be determined. Second, placement algorithms in order to achieve maximum controllability in the system for a given number of controllers will be introduced. 

The following methods focus mostly on HVDC lines. However, with small modifications they can also be applied to Thyristor-Controlled Series Capacitors (TCSCs) and Phase-Shifting Transformers (PSTs), as we will discuss later. In the text, we will use the terms ``HVDC'' and ``controllers'' interchangeably.

\section{Definition of Controllability}
\label{ctrl_sec:def_controllability}
In this work, we aim at maximizing the controllability of the system for a given number of controllable elements. We define controllability as the magnitude of the change in line flows caused by a marginal change in the controller setpoint. In order to maximize controllability in the system we are seeking controller placements that will have the largest impact on line flows and require the smallest control action.

In mathematical terms, for a single controller $j$ we define controllability $H_{c(j)}^{p}$ as the sum of the maximum absolute change on each AC line flow $\Delta P_{L(i)}$  for control actions $|\Delta P_{c(j)}| \leq p$, where $i \in \mathcal{L}=\{1,\ldots,n_L\}$:
\begin{equation}
H_{c(j)}^{p}=\sum_{i=1}^{n_L} max|\Delta P_{L(i)} |, \textnormal{ with \ } |\Delta P_{c(j)}| \leq p.
\label{ctrl_equ:def_controllability}
\end{equation}
Based on \eqref{ctrl_equ:def_controllability}, we are trying to maximize $\sum_j H_{c(j)}^{p} $, in order to maximize controllability. If the system is linear (e.g. DC approximation), $H_{c(j)}^{p}$ will be linear in $p$.

\section{The Controllability Vector $CV$}
\label{ctrl_sec:CV_def}
For the derivations in the following sections we will follow the DC approach (i.e., linearized equations for power flows) and use the Power Transfer Distribution Factors (PTDFs). Nevertheless, we can derive very similar relationships using the full AC approach, as shown in \cite{SCH_PSCC_2014}. For that we need to focus on the bus injections currents and the line currents, and make use of the Kirchhoff current law.

The PTDFs are linear sensitivities, which express the effect of bus injections on the line flows, as shown in \eqref{ctrl_equ:Pl_PTDF_Pb}. $\mathbf{PTDF}$ is an $n_L \times n_B$, where $n_L$ is the number of lines and $n_B$ the number of buses \cite{krause_phd}, \cite{Christie_LODF}. 
\begin{equation}
\mathbf{P_{L}} = \mathbf{PTDF} \cdot \mathbf{P_{B}} \label{ctrl_equ:Pl_PTDF_Pb}
\end{equation}
The equations for the HVDC links follow the formulations introduced in \cite{SCH_Powertech_2013_2} and \cite{Vrakopoulou_ISGT2013}. Each HVDC link is modeled as two virtual voltage sources located at the two nodes where the HVDC line is connected. For each HVDC link, we assume one additional variable for the HVDC power flow. Let $P_{DC}\in \mathbb{R}^{n_{DC}}$ represent the vector of power flows on the HVDC lines. For an HVDC link $j$ connected between nodes $m$ and $n$, the balance between the active power injected or withdrawn from the line is maintained by assuming that $P_{inj,m}=-P_{inj,n}=P_{DC(j)}$ (consistent with the DC power flow approach, HVDC line losses are here neglected).

For an arbitrary change $\Delta P_{DC}$ in the power flow of the HVDC line, the following equation will hold:
\begin{equation}
\Delta \mathbf{P_{L}} = \left(\mathbf{PTDF_{m}} - \mathbf{PTDF_{n}}\right) \cdot \Delta P_{DC},
\label{ctrl_equ:PTDF_diff_DC}
\end{equation}
where $\mathbf{PTDF_{m}}$, $\mathbf{PTDF_{n}}$ are $n_L \times 1$ vectors corresponding to the columns $m$ and $n$ of the $\mathbf{PTDF}$ matrix.

Based on \eqref{ctrl_equ:PTDF_diff_DC}, we define the controllability vector $CV$ to represent the effect of a marginal change of the HVDC line flow on each AC line of the system, as follows:
\begin{equation}
\mathbf{CV_{mn}} = \mathbf{PTDF_{m}} - \mathbf{PTDF_{n}}
\label{ctrl_equ:ctrl_vector}
\end{equation}
Each element of the controllability vector represents the change in the power flow of a specific AC line resulting from the change of the HVDC line setpoint. The higher the absolute value of this element is, the larger the influence that the DC line has on this specific AC line. In Section~\ref{ctrl_sec:plac_algorithm_discussion}, we also derive the controllability vector for TCSCs.

There exists one controllability vector for each pair of nodes in the system. As a result, each HVDC placement between two arbitrary nodes corresponds to a unique controllability vector. The controllability vector will become central in our analysis and the design of the placement algorithm in the following sections.

In the following sections we will derive the upper bound for the number of controllable lines in a system. We will distinguish between the placement of controllers in series and in parallel. Series controllers could be TCSC devices, HVDC back-to-back converters, or replacing an existing AC line with an HVDC line. Controllers in parallel are new HVDC lines that either connect two new pair of nodes or are placed in parallel to an existing line.

\section{Maximum number of controllable lines for full controllability: Placement of Series Controllers}

\begin{figure}[!htb]
    \centering
    \includegraphics[width = 1\columnwidth]{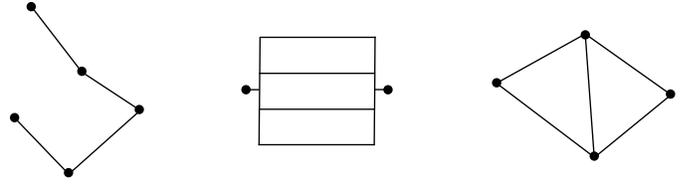}
    \caption{Graph decomposition. Every graph can be decomposed to series (left), parallel (middle), and wheatstone (right) graphs.}\label{ctrl_fig:graph_decomp}\vspace{-0.3cm}
\end{figure}

Power grids have a planar graph topology. It has been shown that any planar graph can be decomposed to a set of series, parallel, and Wheatstone graphs \cite{duffin_graph_decomp}. Figure~\ref{ctrl_fig:graph_decomp} shows the three different graph types. Showing that the following derivations apply to each of these graph types, it follows that it will apply to any power grid topology which is a synthesis of such graphs.

We assume a power system with $n_B$ nodes and $n_L$ branches connecting the nodes. For every node $k$ it will hold:
\begin{equation}
P_k=P_{ki}+P_{kj}+P_{km}+\ldots
\label{ctrl_equ:fully_control_single}
\end{equation}
where $i,j,k,m,\ldots \in \{1,\ldots, n_B\}$. $P_{ki}, P_{kj}, \ldots$ represent the line flows on lines connected at node $k$, while $P_k$ represents the net power injections (resp. withdrawals) at the same node. If no shunt element (e.g. generator, load, shunt capacitance, etc.) is connected to a node $k$, then $P_k=0$.

We now assume that the system is fully controllable. This means that given a certain profile of power injections at all nodes, we can determine independently the power flows on each line. Such a system could be thought of as a network of HVDC point-to-point links. 

The algebraic system of \eqref{ctrl_equ:fully_control_single} $\forall k \in n_B$ is of the form $\mathbf{A}\mathbf{x}=\mathbf{B}$:
\begin{equation}
\mathbf{P_B}=\mathbf{A}\mathbf{P_L}
\label{ctrl_equ:AxB}
\end{equation}
$\mathbf{A}$ is a $n_B \times n_L$ matrix whose elements are ones or zeros. $\mathbf{P_B}$ is a $n_B \times 1$ vector equal to the bus power injections: $\mathbf{P_B}=[P_1 \ldots P_{n_B}]^T$. $\mathbf{P_L}$ corresponds to a $n_L \times 1$ vector containing all the line flows, i.e. $\mathbf{P_L}=[P_{ij}]$, with $i,j \in \{1,\ldots,n_B\}$.

We assume the net injections/withdrawals are given, so $\mathbf{P_B}$ is a vector of constants. The question is how many elements of the $\mathbf{P_L}$ vector should be known (or ``exogenously constrained'') in order to be able to fully define the power flows, i.e. the rest elements of the $\mathbf{P_L}$ vector.

For a system of equations to have a unique solution, matrix $\mathbf{A}$ must be square and non-singular. In a meshed power system, however, there are usually more branches than nodes, i.e. $n_L>n_B$. This means that our problem is underdetermined.

Assuming no losses in the lines, for all nodal power injections $P_k$ must hold $\sum_k P_k =0$. This implies that the rows of $\mathbf{A}$ in \eqref{ctrl_equ:AxB} are linearly dependent. In total, we have $n_B-1$ linearly independent equations. We assume here that all parallel lines between two nodes are represented by one equivalent line. Thus, matrix $A$ is of full column rank.

If $n_L < n_B$, it must hold $n_L=n_B-1$, since each branch connects two nodes. Having also $n_B-1$ linearly independent rows, matrix A is square and non-singular, and our problem yields one unique solution. Consequently, for a given set of  bus injections, if $n_L=n_B-1$ (e.g. see series graph in Fig.~\ref{ctrl_fig:graph_decomp}) none of the line flows can be controlled to an arbitrary setpoint.

If $n_L=n_B$ and we have $n_B-1$ linearly independent equations, we need to define one more equation in order to have a unique solution of the system. As a result, if $n_L=n_B$ we need one controller to set one line flow to a specific value. According to this value the rest of the line flows can be calculated.

If $n_L > n_B$, then we can freely select $n_L-n_B+1$ variables. As a result, even in a fully controllable power system we can freely determine the flow only on $n_L-n_B+1$ lines. The rest of the line flows will be determined through the bus power injections (see also the parallel and Wheatstone graphs in Fig.~\ref{ctrl_fig:graph_decomp}). This implies that during a controller placement process -- converting an AC non-controllable system to a system with controllable power flows -- we would not need to place more than $n_L-n_B+1$ series controllers to control the maximum number of lines for a given set of bus power injections.

This should hold, as long as the controllers are assumed to have infinite (i.e. sufficiently large) boundaries and there is sufficient transmission capacity. If their operation is limited, either we might need more controllers, or we might never achieve full controllability of the power flows. This leads us to the following two conclusions, as shown in \eqref{ctrl_equ:min_num_ctrl_for_full_ctrl_series}. First, by placing controllers in series with existing AC lines, we cannot control more than $n_L-n_B+1$ branches. Second, in order to achieve full controllability we need at least $n_L-n_B+1$ series controllers:
\begin{equation}
sup N_{\textnormal{ctrl.lines}}^{\textnormal{series}} = n_L-n_B+1 = inf \, N_\textnormal{controllers}^{\textnormal{series}}.
\label{ctrl_equ:min_num_ctrl_for_full_ctrl_series}
\end{equation}

\section{Maximum number of controllable lines for full controllability: Placement of Controllers in Parallel}
\label{sec:max_controllable_lines}
According to control systems theory, full controllability describes the ability of an external input to move the internal state of a system from any initial state to any other final state in a finite time interval. We can determine how many states of the system are controllable from the controllability matrix. If the controllability matrix has full row rank, then the system is controllable. We define the discrete-time invariant system\footnote{The results would be similar for a continuous LTI system. Then $\dot{\mathbf{x}}=\mathbf{0}\mathbf{x}(t)+B\mathbf{u}(t)$. The controllability matrix is $\mathbf{C}=[\mathbf{B}]=[\mathbf{CV}]$, and $rank(C)=n_B-1$ }, as shown in \eqref{equ:CV_LTI}. 
\begin{equation}
\mathbf{P_{L}}(k+1) = \mathbf{I} \cdot \mathbf{P_{L}}(k) + \mathbf{CV} \cdot \mathbf{P_{DC}}(k) \label{equ:CV_LTI}
\end{equation}
Then the controllability matrix $\mathbf{C}$ is equal to:
\begin{equation}
\mathbf{C}=[\mathbf{CV} \,\, \mathbf{I} \cdot \mathbf{CV} \,\, \ldots \,\, \mathbf{I}^{n_{L}-1}\mathbf{CV}]
\label{equ:Ctrl_CV_LTI}
\end{equation}
From \eqref{equ:Ctrl_CV_LTI}, it is $rank(\mathbf{C})=rank(\mathbf{CV})$. However, $\mathbf{CV}$ is only a linear transformation of $\mathbf{PTDF}$, i.e. $\mathbf{CV}=\mathbf{T}\cdot \mathbf{PTDF},$ where $\mathbf{T}$ has dimensions $n_L \times n_{DC}$ and is of full rank, since it is a set of linearly independent vectors containing $0, 1, -1$. Assuming that there are controllers installed to control flows between every possible pair of nodes, then $n_{DC}= n_B(n_B-1)/2 > n_B$ and $rank(\mathbf{CV})=rank(\mathbf{PTDF})= n_{B}-1$.

This means that for a given set of power injections, we can control up to $n_B-1$ lines by installing parallel power flow controllers, see \eqref{ctrl_equ:min_num_ctrl_for_full_ctrl_parallel}.  For example, in the series graph of Fig.~\ref{ctrl_fig:graph_decomp}, we can install up to four parallel controllers and we will be able to control different degrees of freedom of the system.
\begin{equation}
sup N_{\textnormal{ctrl.lines}}^{\textnormal{parallel}} = n_B-1 = inf \, N_\textnormal{controllers}^{\textnormal{parallel}}.\label{ctrl_equ:min_num_ctrl_for_full_ctrl_parallel}
\end{equation}
As we will see in the following sections, each controller (either DC lines or TCSC) can only control a single line. As a result, even in the ideal case where all line flows were fully controllable and the controllers had infinite bounds, at any given point no more than $n_B-1$ controllers could be active.

\section{Adding a controller to the system}
\label{ctrl_sec:adding_controller}
In this section we now assume that we have a non-controllable AC network. We further assume that the line capacities are sufficiently high so that no congestion is induced by different bus power injections. We will focus on two main power flow control elements: the HVDC line and the Thyristor-Controlled Series Capacitor (TCSC). We will show that in the general case each of these elements can determine the flow on almost any single AC line, but only one at a time. 
\subsection{HVDC line}
From \eqref{ctrl_equ:PTDF_diff_DC}~and~\eqref{ctrl_equ:ctrl_vector}, it follows that:
\begin{equation}
\mathbf{P_L} = \mathbf{CV_{mn}} \cdot P_{mn}^{DC},
\label{ctrl_equ:HVDC_control}
\end{equation}
where $m,n$ are the nodes connected through the HVDC line. With $P_{mn}^{DC}$ being a scalar, there is only one degree of freedom, and therefore each HVDC line can control only a single AC line at a time. Through \eqref{ctrl_equ:HVDC_control}, we observe that by adding an HVDC line, we can influence all AC power flows in the system as long as all elements of the vector $C_{mn}$ are non-zero. This is most often the case. In a meshed transmission system it is relatively rare that power injection and withdrawal between a pair of nodes $m,n$ influences in exactly the same way a line flow $k$ (so that $PTDF_{m,k}$, $PTDF_{m,k}$, or $CV_{mn,k}$ equal zero). As a result, by adding one HVDC line we can control the flow of almost any single AC line.


\subsection{Thyristor-Controlled Series Capacitor (TCSC)}
A TCSC is connected in series to a transmission line and is able to control the reactance of this line. As a result, it can indirectly control the line flow either on this line, or influence the flow of any other line in the system. Through the following derivations we show that one TCSC device can influence almost all line flows in the system (to different degrees) and can arbitrarily determine the line flow on exactly one line. 

In a DC power flow context it holds:
\begin{equation}
\mathbf{P_L}= \mathbf{PTDF} \cdot \mathbf{P_B} \Leftrightarrow \mathbf{P_L}=\mathbf{B_L} \mathbf{\tilde{B}_B}^{-1} \mathbf{P_B}
\end{equation}
$\mathbf{B_L}$ is the line susceptance matrix. $\mathbf{\tilde{B}_B}^{-1}$ is the inverse of the bus susceptance matrix $\mathbf{B_B}$\footnote{Because the bus susceptance matrix is singular, before inverting it we eliminate the row and the column corresponding to the slack bus. After the inversion we insert zero vectors at this row and column which correspond to the slack bus. To denote this manipulation, we express the inverse of $\mathbf{B_B}$ with a tilde, i.e. $\mathbf{\tilde{B}_B}^{-1}$.}.

We wish to examine the change in the line flows $\mathbf{P_L}$, by changing the line reactance $x_{ij}$. It is:
\begin{align}
\frac{\partial \mathbf{P_L} }{\partial x_{ij}} &= \frac{\partial(\mathbf{B_L} \mathbf{\tilde{B}_B}^{-1})}{\partial x_{ij}} \mathbf{P_B} + \mathbf{B_L} \mathbf{\tilde{B}_B}^{-1} \underbrace{\frac{\partial \mathbf{P_B}}{\partial x_{ij}}}_{=0}\\
\frac{\partial \mathbf{P_L} }{\partial x_{ij}} &= \frac{\partial(\mathbf{B_L})}{\partial x_{ij}}\mathbf{\tilde{B}_B}^{-1} \mathbf{P_B} + \mathbf{B_L} \mathbf{\tilde{B}_B}^{-1} \frac{\partial(\mathbf{\tilde{B}_B})}{\partial x_{ij}} \mathbf{\tilde{B}_B}^{-1} \mathbf{P_B} \\
\frac{\partial \mathbf{P_L} }{\partial x_{ij}} &= \frac{\partial(\mathbf{B_L})}{\partial x_{ij}}\mathbf{\tilde{B}_B}^{-1} \mathbf{P_b} + \mathbf{PTDF} \frac{\partial(\mathbf{\tilde{B}_B})}{\partial x_{ij}} \mathbf{\tilde{B}_B}^{-1} \mathbf{P_B} \label{ctrl_equ:TCSC_thetaPl}
\end{align}
From \eqref{ctrl_equ:TCSC_thetaPl}, the term $\mathbf{PTDF} \frac{\partial(\mathbf{\tilde{B}_B})}{\partial x_{ij}} \mathbf{\tilde{B}_B}^{-1}$ will always result in a non-sparse matrix (in the general case where no radial connections exist all values will be non-zero, except for the column corresponding to the slack bus). This leads to the conclusion that in a meshed system a TCSC can influence almost all AC lines in the system. Given that one TCSC has a single degree of freedom, assuming that it has sufficiently large boundaries, it can determine arbitrarily the flow on any single line in the system. 

\section{Example for Placement of Series Controllers -- 10-bus network}
Consider the power system model of Fig.~\ref{ctrl_fig:swiss_powersystem}. It consists of 10 nodes and 14 AC lines. The system data can be found in the Appendix of~\cite{SCH_PESGM_2011}.
\begin{figure}[htb]
    \centering
    \includegraphics[width = 0.75\columnwidth]{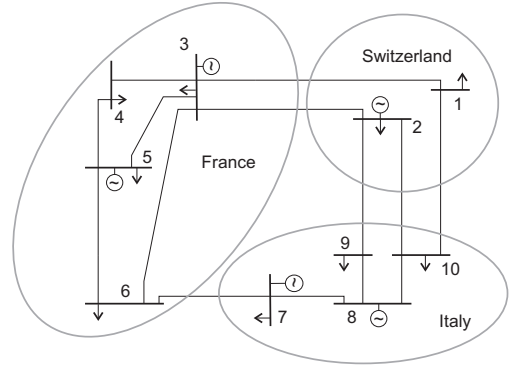}
    \vspace{-0.3cm}
    \caption{10-bus network used for the simulations.}\label{ctrl_fig:swiss_powersystem}\vspace{-0.3cm}
\end{figure}
In this example we will examine the placement of TCSC devices. We will show that, given a single generation and demand snapshot, there is a maximum number of controllable devices we can install in a power system, above which no additional power flow controllability can be gained. This upper bound is equal to $n_L-n_B+1$. For this specific case study, with $n_B=10$ and $n_L=14$, this amounts exactly to 5 TCSC devices.

We start by having no TCSC device installed in our system. We run a Standard AC-OPF and a Security-Constrained OPF and compute the Cost of Security -- as defined in \eqref{ch3_equ:CoS_def} -- in percentage values. For the SC-OPF algorithm, we use the formulation described in \cite{SCH_PESGM_2011}. Subsequently, we add a TCSC device along an existing line and compute the Cost of Security (CoS) again. We do that for each possible line placement, before we move on to the installation of two TCSC devices at the same time. All possible combinations of line pairs are examined. 

\begin{figure}[htb]
\scriptsize
\psfrag{Gen. Costs Difference: AC-OPF vs SC-OPF (in \%)}[c][c]{Cost of Security (in \%)}
\psfrag{TCSC}[c][c]{Number of TCSC devices}
\psfrag{0}[r][c]{0}
\psfrag{1}[r][c]{1}
\psfrag{2}[r][c]{2}
\psfrag{3}[r][c]{3}
\psfrag{4}[r][c]{4}
\psfrag{5}[r][c]{5}
\psfrag{6}[r][c]{6}
\psfrag{7}[r][c]{7}
\psfrag{8}[r][c]{8}
\psfrag{9}[r][c]{9}
\psfrag{10}[r][c]{10}
\psfrag{0Sr}[c][c]{0}
\psfrag{1Sr}[c][c]{1}
\psfrag{2Sr}[c][c]{2}
\psfrag{3Sr}[c][c]{3}
\psfrag{4Sr}[c][c]{4}
\psfrag{5Sr}[c][c]{5}
\psfrag{6Sr}[c][c]{6}
\psfrag{7Sr}[c][c]{7}
\psfrag{8Sr}[c][c]{8}
\psfrag{9Sr}[c][c]{9}
\psfrag{10Sr}[c][c]{10}
\psfrag{11Sr}[c][c]{11}
\psfrag{12Sr}[c][c]{12}
\psfrag{13Sr}[c][c]{13}
\psfrag{14Sr}[c][c]{14}
    \centering
    \includegraphics[width=1\columnwidth]{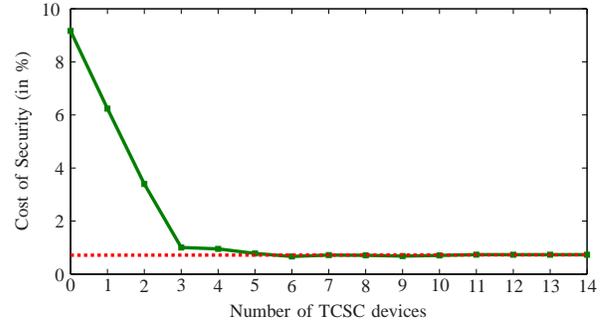}\vspace{-0.3cm}
    \caption{Cost of Security (in \% relative to AC-OPF Costs) for different number of TCSC devices installed}\label{ctrl_fig:SCOPFvsACOPF_percent}
\end{figure}

Figure~\ref{ctrl_fig:SCOPFvsACOPF_percent} presents the Cost of Security in percentage, after placing one TCSC, two TCSCs, etc. up to 14 TCSC devices. The values shown are for the TCSC placement that achieved the highest CoS reduction from all possible combinations at each placement step.
With the installation of TCSC devices both the AC-OPF and the SC-OPF costs decrease. At the same time, their difference, both in absolute and percentage terms, also decreases. As shown in Fig.~\ref{ctrl_fig:SCOPFvsACOPF_percent}, after the placement of the first TCSC device we observe a significant reduction in the CoS. The controllability introduced through this TCSC offers additional flexibility in the routing of the power flows which reduces the amount of redispatching necessary to maintain the system N-1 secure. It is interesting to note that after the placement of five TCSC devices, the CoS stabilizes and no further significant reduction on the CoS can be observed, e.g. by adding six or more TCSCs.

According to our considerations, the minimum number of TCSC devices for full controllability is 5 controllers ($n_L=14$, $n_B=10$, see \eqref{ctrl_equ:min_num_ctrl_for_full_ctrl_series}). This is confirmed from the simulation results shown in Fig.~\ref{ctrl_fig:SCOPFvsACOPF_percent}. The placement of further controllers does not add anything to the controllability of this system, and it has no effect for the cost reduction of the specific snapshot.

Two additional comments about this case study should be made here. First, as we can observe, the CoS is not eliminated but amounts to about 0.72\% of the AC-OPF costs after the placement of five or more TCSCs. Second, the CoS ``stabilizes'' at about 0.72\% but small oscillations around this value can be observed with the placement of additional TCSC devices (i.e. above five). There are two reasons for these two phenomena. First, in order to completely eliminate the Cost of Security not only controllability plays a role but also the transmission capacity. In case of contingency, enough transmission capacity should be available, so that the power could be rerouted through alternative paths. This means that in case of higher loading in the system the CoS could ``stabilize'' at values higher than 0.72\%, while for lower system loading CoS can equal zero. The second reason has to do with the TCSC limits. For this case study, we have selected realistic limits for the control capabilities TCSC devices: the TCSC can vary the line reactance up to 60\%. With sufficiently large limits, the CoS could be completely ``stabilized'', and eliminated if there is sufficient transmission capacity. An additional reason for these small ``oscillations'' is the fact that both the AC-OPF and the SC-OPF are non-linear and non-convex problems. As a result, there might be small differences due to numerical reasons, as the solver might not be able to find the absolute optimal solution.

\section{HVDC Placement for Maximum Controllability: Placement of Parallel Controllers}
In the rest of this paper, we propose algorithms for the placement of HVDC lines to maximize the controllability in the system, based on two different definitions of controllability. The goal is to achieve the maximum controllability in the system with the minimum number of controller placements.

The first algorithm introduced in this paper makes extensive use of the controllability vector $CV$. Dealing with vectors for the placement procedure allows a better ``visualization'' of the process and, hopefully, allows a more intuitive way to explain the obtained results. Further, taking advantage of linear algebra properties and metrics, such as orthogonality and norms, we are able to save significantly on computational effort, thus reducing the computational time. In the following sections we outline the algorithm based on definition \eqref{ctrl_equ:def_controllability}.

\subsection{Placement Criteria}
The placement of the controller should be based on two main criteria. First, the influence of the controller on a specific line. Seeking to maximize controllability as defined in~\eqref{ctrl_equ:def_controllability}, the most effective controller placement is at a location where a marginal change in the controller setpoint will result in the maximum change of the line flow. The second criterion is the number of individual lines it can control in an effective manner. 
In the ideal case, we are seeking a controller placement which can control as many lines as possible, and, at the same time, it will need the least control actions to achieve a power flow change. As these two objectives cannot always be satisfied at the same time, we seek the best possible compromise. We will see in the following how we translate these two objectives in equivalent metrics, and how we design our algorithm in order to identify the optimal controller placement.

\subsection{Metrics for the placement criteria}
\label{ctrl_sec:metrics_placement_criteria}
Our approach is based on the controllability vector $CV$ introduced in Section~\ref{ctrl_sec:CV_def}. A suitable placement metric would be the 1-norm of the controllability vector $CV$, for two reasons: first, it takes into account all vector elements (in comparison with the $\infty$-norm), and second it handles uniformly all vector elements, i.e. it does not give an advantage/penalty to extreme values, such as the L2 and higher order norms.

The placement of the second controller should influence a set of lines as different as possible from the first controller: we are thus seeking a $CV$ orthogonal to the first one. Still, the vector norm should be taken into account, as this determines the degree of influence. In this paper, we select as metric the volume of the polytope defined by the vectors in question. 

\begin{figure}[htb]
    \centering
    \includegraphics[width=0.8\linewidth]{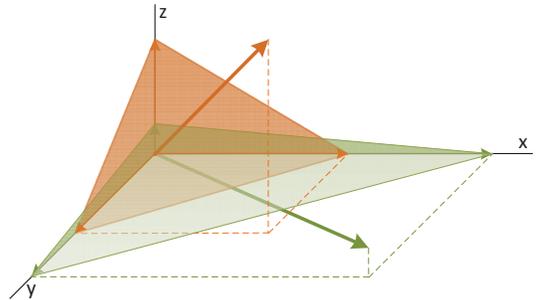}
    \caption{Schematic representation of the 1-norm and the convex hull volume metrics. The orange controllability vector has a lower 1-norm but its convex hull covers a larger volume than the green vector. }\label{ctrl_fig:norm1_vs_volume_schematic}
\end{figure}

Here, we will focus on the motivation for using the $CV$ vectors and the volume as placement metric. We define the polytope from the projections of the vectors on the coordinate axes, as explained in the following. In general, the larger the volume covered by the polytope, the higher and more diverse is the influence of the controllers on the line flows. This is schematically presented in Fig.~\ref{ctrl_fig:norm1_vs_volume_schematic}. Assume that axes x, y, z represent the flows on three AC lines. The orange and green vectors are controllability vectors, representing the influence of a controller placement on these lines. The higher the projection of the vector on each axis is, the higher the effect that a marginal change of this controller has on the specific line flow. The convex hull of each CV is determined from the projection of the CV on each axis. As we can observe the green vector has a substantial influence on two of the lines (along x and y axes), but it does not affect significantly the third line (along the z axis). On the other hand, the orange controllability vector has a more balanced influence on all lines. Selecting the 1-norm as a metric, the green vector has a higher 1-norm than the orange vector. On the other hand, the convex hull resulting from the orange vector results in a larger volume than the convex hull from the green vector and has a more uniform effect on a larger set of lines. In Section~\ref{ctrl_sec:plac_algorithm_case_study} we compare the two placement metrics. 

To calculate the convex hull, we transform the vector $CV=[c_i]$ into the matrix $X=diag([c_i])$, where each row corresponds to a point in the $n_L$-dimension space. We augment matrix $X$ by adding the point of origin, i.e. a zero vector, so that $X'=diag([0\,c_i])$. The set of these points define a conical hull, which is by definition convex. Convexity is a necessary condition in order to compute the resulting volume. 

As shown in~\eqref{ctrl_equ:conichull}, the conical hull is the set of all conical combinations for a given set S.
\begin{equation}
coni(S) = \left\{ \sum_{i=1}^k \alpha_i x_i | x_i \in S, \alpha_i \in \mathbb{R}, \alpha_i \geq 0, i, k =1,2, \ldots \right\} 
\label{ctrl_equ:conichull}
\end{equation}
Based on \eqref{ctrl_equ:PTDF_diff_DC}, \eqref{ctrl_equ:ctrl_vector}, \eqref{ctrl_equ:conichull}, vectors $x_i$ are the row vectors of matrix $X'$, while the scalars $a_i$ represent $\Delta P_{DC}$. Contrary to this definition, $\Delta P_{DC}$ could also be negative. In order to account for that in our algorithm, we consider instead both $CV$ and $-CV$ for each placement. A detailed description of the algorithm follows in the next sections.

\subsection{$CV$-based Placement Algorithm}
\label{sec:vectorbasedalg}
\begin{figure}[htb]
    \centering
    \includegraphics[width=0.75\linewidth]{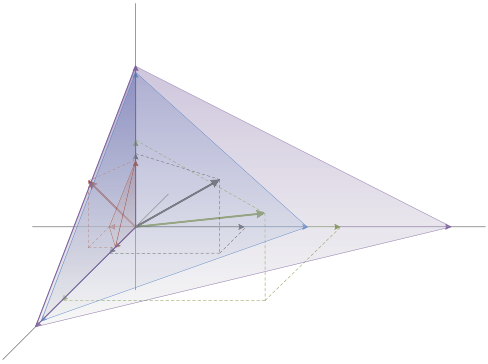}
    \caption{Schematic representation of the convex hull volumes resulting from two nearly orthogonal vectors (green and red; their addition is shown in blue) in comparison with the volume of the convex hulls resulting from vectors not too orthogonal (green and black; their addition is shown in purple). }\label{ctrl_fig:volumes_orthogonal_schematic}
\end{figure}

In order to have a fair comparison among all controllers, we assume the same marginal change in the controller flow for all possible placements: $\Delta P_{DC}=1$. 
As placement measure we use the sum of these volumes. The larger the sum, the more effective the controller placement. For each subsequent placement, in order to avoid adding overlapping volumes, we determine the extreme values for each orthant and calculate the resulting volumes. To reduce the number of subsequent placements to be examined, we consider $CV$ which are as orthogonal as possible to the vectors we have already selected \footnote{By definition, two vectors can either be orthogonal or not. Here, by the phrase ``as orthogonal as possible'' we imply that the angle between the two vectors should be as close to $90^{o}$ as possible. For the sake of readability in this and the following sections, we will abuse the term ``orthogonal'' in order to compare the angles between vectors, e.g. ``less orthogonal'' will denote that the angle between a pair of vectors deviates further away from $90^{o}$.}. As shown in Figure~\ref{ctrl_fig:volumes_orthogonal_schematic}, such vectors would not necessarily result in the largest possible convex hull volume, but they will offer more diverse controllability options. 

The algorithmic steps are outlined below: 
\begin{enumerate}
\item Form diagonal matrices from the $CV$, including the point of origin: $\mathbf{X'_{mn}}=[\mathbf{0}; diag([c^{mn}_i])]$
\item Calculate $V^{mn}_{1}=vol(\mathbf{X'_{mn}})+vol(-\mathbf{X'_{mn}})$
\item Place the first controller between the nodes $m,n$ that resulted in the largest $V^{mn}_{1}$.
\item Calculate the orthogonal projections of all other controllability vectors $\mathbf{CV_{ij}}$ on the vector space defined by $\mathbf{CV_{mn}}$.\label{iter_step}
    \begin{enumerate}
\item Calculate matrix: $\mathbf{P_A}=\mathbf{A}(\mathbf{A}^{T}\mathbf{A})^{-1}\mathbf{A}^{T}$, where $\mathbf{A}=[\mathbf{CV_{mn}}]$
\item Calculate the orthogonal projections: $\mathbf{CV^{proj}_{ij}}=\mathbf{P_A} \mathbf{CV_{ij}}$, for $i,j \neq m,n$
\end{enumerate}
\item Calculate: $cos\phi_{ij}=\frac{\parallel \mathbf{CV^{proj}_{ij}}\parallel}{\parallel \mathbf{CV_{ij}}\parallel}$. It holds $cos\phi_{ij} \in [0,1]$.
\item Select the $k$-vectors $CV_{ij}$ with the lowest $cos\phi_{ij}$.
\item For all $k$-vectors formulate matrix $\mathbf{S}=[CV_1\quad$ \mbox{$-CV_1$} $\quad CV_2\quad$ \mbox{$-CV_2$}$\quad$ \mbox{$CV_1+CV_2$}$\quad$ \mbox{$CV_1-CV_2$}$\quad$ \mbox{$-CV_1+CV_2$}$\quad$ \mbox{$-CV_1-CV_2$}$]$.
\item Divide the column vectors of matrix $\mathbf{S}$ in the orthants they belong to. For each orthant $p$ form a matrix $S^{orth,p}$ with the respective column vectors.
\item Select the extreme values from each row vector of matrix $S^{orth,p}$ and form the vector $S^{orth-max,p}$.\label{orthant_iter_step_1} (i.e. in order to avoid adding overlapping volumes).
\item Compute the volume  $V_{2,p}^{ij}$ of the conical hull defined from the row vectors of matrix $S^{orth-max-vol,p}=diag([0; \, s^{orth-max,p}_i])$. \label{orthant_iter_step_2}
\item Repeat Steps~\ref{orthant_iter_step_1}~and~\ref{orthant_iter_step_2} for all $p$ orthants.
\item Compute the sum of the conical hull volumes for this placement: $V_{2}^{ij} = \sum_p V_{2,p}^{ij}$.
\item For the second placement, select the nodes $i,j$, for which $V_2$ becomes maximum.
\item For subsequent placements go to Step~\ref{iter_step}, now assuming the vector space will be defined by the vectors $A=[CV_{mn}~CV_{ij} \ldots]$
\end{enumerate}

\subsection{Case Study}
\label{ctrl_sec:plac_algorithm_case_study}

\paragraph{Comparison of the two metrics: 1-norm vs. volume of convex hull}
Consider the power system model of Fig.~\ref{ctrl_fig:swiss_powersystem}. In this paragraph we compare the 1-norm metric with the volume of the convex hull defined by each $CV$. We used function \verb|convhulln| of \textsc{Matlab} for the volume calculation. 
\begin{figure}[!htb]
\scriptsize
\psfrag{0}[r][c]{0}
\psfrag{0.5}[r][c]{0.5}
\psfrag{1.0}[r][c]{1.0}
\psfrag{1.5}[r][c]{1.5}
\psfrag{2.0}[r][c]{2.0}
\psfrag{2.5}[r][c]{2.5}
\psfrag{3.0}[r][c]{3.0}
\psfrag{3.5}[r][c]{3.5}
\psfrag{4.0}[r][c]{4.0}
\psfrag{4.5}[r][c]{4.5}
\psfrag{1-normCV}[c][c]{1-norm of $CV$}

    \centering
    \includegraphics[width=1\linewidth]{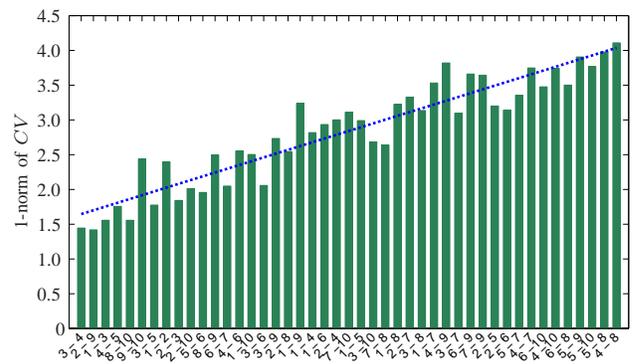}
    \caption{1-norm of Controllability Vectors $CV_{mn}$. The vectors are sorted from minimum to maximum volume of the convex hull they define. The blue dashed line represents the trendline, denoting a correlation between the convex hull volumes and $\parallel CV_{mn} \parallel_1$ }\label{ctrl_fig:1normCV_vs_vlmcvxhull_1}\vspace{-0.3cm}
\end{figure}

\begin{figure}[!htb]
\scriptsize
\psfrag{0d0}[r][c]{0}
\psfrag{0d5}[r][c]{0.5}
\psfrag{1d0}[r][c]{1.0}
\psfrag{1d5}[r][c]{1.5}
\psfrag{2d0}[r][c]{2.0}
\psfrag{2d5}[r][c]{2.5}
\psfrag{3d0}[r][c]{3.0}
\psfrag{3d5}[r][c]{3.5}
\psfrag{4d0}[r][c]{4.0}
\psfrag{1-normorthogvector}[c][t]{1-norm of $CV$}
\psfrag{Placement of Second HVDC line}[c][c]{Placement of Second HVDC line}
    \centering
    \includegraphics[width=1\linewidth]{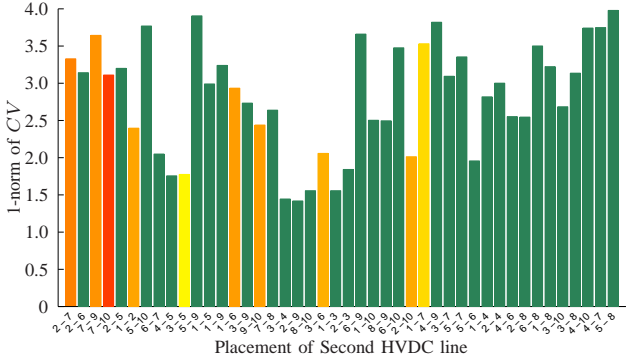}
    \caption{Second HVDC placement. First HVDC placed between 4-8. The CVs are sorted with increasing volume of the convex hull they create in combination with $CV_{48}$. Red/yellow-hued bars denote the most orthogonal CVs, i.e $cos \phi_{(CV_{ij}-CV_{48})} \leq 0.2$. Red are closest to zero, yellow closest to 0.2. The CV to be selected is one with $cos \phi \leq 0.2$ that results in the largest volume. Here, this is between nodes 1-7.} \label{ctrl_fig:plac_algor_2nd_placement}
\end{figure}

Figure~\ref{ctrl_fig:1normCV_vs_vlmcvxhull_1} presents the 1-norm of all controllability vectors, \emph{sorted from minimum to maximum volume of the convex hulls they define}. The blue dashed line shows the trendline of the 1-norm vectors -- it is evident that there is a correlation between the two metrics. Pair 4-8 results in the highest ``controllability'' with both metrics. The bottom-five placements are the same for both metrics, nevertheless with a slightly different ranking. For the reasons outlined above, the volume of convex hull seems to represent better the degrees of freedom that are added through the placement of each controller.
\paragraph{Placement of three HVDC lines}
As shown from the results presented in Fig.~\ref{ctrl_fig:1normCV_vs_vlmcvxhull_1}, the first HVDC line should be placed between nodes 4 and 8. Figure~\ref{ctrl_fig:plac_algor_2nd_placement} sorts the 1-norm vectors based on the polytope volume for the second placement. The $CV_{ij}$ ``most orthogonal'' to $CV_{48}$, i.e. with $cos\phi \leq 0.2$ are colored red to yellow. The $CV_{17}$ is selected for the second placement.
We repeat the same procedure for the third HVDC line placement (figure not shown due to space limitations). In this case, only two vectors are below the ``orthogonality'' limit, $CV_{9-10}$ and $CV_{2-10}$, thus the algorithm saves significantly on computation time. The third line is placed between nodes 9 and 10.

In the following section, we compare these results with an optimal placement algorithm we developed based on an alternative definition for controllability.
\section{Optimal Placement based on an Alternative Definition of Controllability}
A definition for controllability alternative to \eqref{ctrl_equ:def_controllability} is the following: we define controllability $H_{c}^{p}$ as the sum of control actions $|\Delta P_{c(j)}|$ for changes in line flows $|\Delta P_L(i)| \leq p$, on all AC lines $i \in \mathcal{L}=\{1,\ldots,n_L\}$:
\begin{equation}
H_{c}^{p}=\sum_{j=1}^{n_c} min|\Delta P_{c(j)} |,
\label{ctrl_equ:def_controllability_alternative}
\end{equation}
so that $\forall i \in \mathcal{L}$, $|\Delta P_{L(i)}|$ can take \emph{any} value between $\pm p$. Assuming the DC approximation for the power flow, $H_{c}^{p}$ will be linear in $p$. Compared with \eqref{ctrl_equ:def_controllability}, here we define controllability as the \emph{magnitude of the control action} necessary to perform a change in each of the line flows, instead of the \emph{magnitude of the change in line flows} for a marginal controller change; so here we are looking for the minimum $H_{c}^{p}$. Based on \eqref{ctrl_equ:def_controllability_alternative} and vector $CV$, we formulate the following optimization problem. For every different HVDC placement and each ``set'' of AC line flows:
\begin{equation}
\min \mathbf{1}^{T} \mathbf{t}
\label{ctrl_equ:exhaust_search_obj_fun}
\end{equation}
subject to:
\begin{equation}
\begin{aligned}[l]
-\mathbf{t} &\leq  \mathbf{P^{DC}} \leq \mathbf{t}\\
&\mathbf{0} \leq \mathbf{t} \\
\end{aligned}
\qquad 
\begin{aligned}[r]
\mathbf{CV} &\cdot \mathbf{P^{DC}} = \Delta \mathbf{P_{L}}\\
-\mathbf{P^{DC}_{max}} & \leq \mathbf{P^{DC}} \leq \mathbf{P^{DC}_{max}}
\end{aligned}
\end{equation}
By the term ``set'', we denote the set of AC lines to be independently controlled during each HVDC placement. For example, two HVDC lines can independently control two AC lines, so for the second HVDC placement we run the algorithm iteratively for all pairs of AC lines and we add the magnitude of control actions. We test three different ways to determine $\Delta \mathbf{P_{L}}$. First, constant $\Delta \mathbf{P_{L(i)}} = 100 MW$ for all $i \in n_L$. Second, $\Delta \mathbf{P_{L(i)}}$ proportional to the line limit of line $i$ ($\Delta \mathbf{P_{L(i)}} = 0.1 F_{L(i)}$). Third, $\Delta \mathbf{P_{L(i)}}$ proportional to the line reactance of line $i$ ($\Delta \mathbf{P_{L(i)}} = 1000 \cdot x_{L(i)}$), with $x_{L{i}}$ in p.u. 
\begin{figure}[htb]
\scriptsize
\psfrag{0}[r][c]{0}
\psfrag{10}[r][c]{10}
\psfrag{20}[r][c]{20}
\psfrag{30}[r][c]{30}
\psfrag{40}[r][c]{40}
\psfrag{50}[r][c]{50}
\psfrag{60}[r][c]{60}
\psfrag{70}[r][c]{70}
\psfrag{80}[r][c]{80}
\psfrag{90}[r][c]{90}
\psfrag{100}[r][c]{100}
\psfrag{Relative Magnitude of Control Actions (in \%)}[c][t]{Relative Magnitude of Control Actions (in \%)}
\psfrag{Placement of Second HVDC line}[c][c]{Placement of Second HVDC line}
    \centering
    \includegraphics[width=1\linewidth]{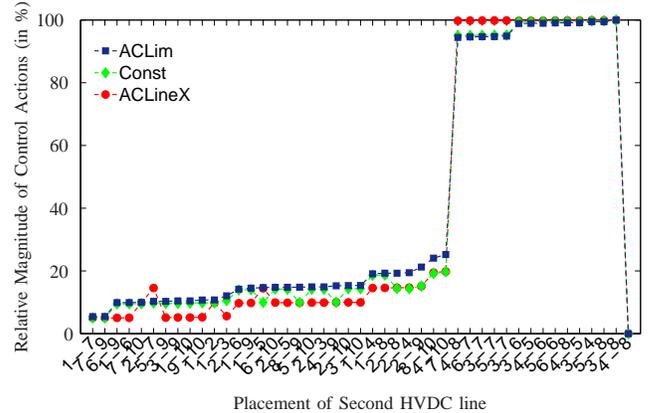}
    \caption{Second HVDC line -- alternative controllability definition. Sum of control actions for each placement (expressed in \% relative to the worst case). [ACLim]: $\Delta \mathbf{P_{L(i)}}$ proportional to the line limit; [Const]: $\Delta \mathbf{P_{L}}$ the same for all lines; [ACLineX]: $\Delta \mathbf{P_{L}}$ proportional to line reactance.} \label{ctrl_fig:series_plac_2line}\vspace{-0.3cm}
\end{figure}\begin{figure}[htb]
\scriptsize
\psfrag{0}[r][c]{0}
\psfrag{10}[r][c]{10}
\psfrag{20}[r][c]{20}
\psfrag{30}[r][c]{30}
\psfrag{40}[r][c]{40}
\psfrag{50}[r][c]{50}
\psfrag{60}[r][c]{60}
\psfrag{70}[r][c]{70}
\psfrag{80}[r][c]{80}
\psfrag{90}[r][c]{90}
\psfrag{100}[r][c]{100}
\psfrag{Relative Magnitude of Control Actions (in \%)}[c][t]{Relative Magnitude of Control Actions (in \%)}
\psfrag{Placement of Second HVDC line}[c][c]{Placement of Third HVDC line}
    \centering
    \includegraphics[width=1\linewidth]{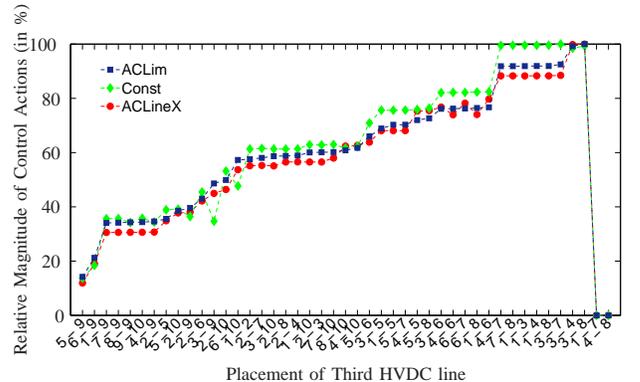}
    \caption{Third HVDC line -- alternative controllability definition. Sum of control actions for each placement (expressed in \% relative to the worst case). [ACLim]: $\Delta \mathbf{P_{L(i)}}$ proportional to the line limit; [Const]: $\Delta \mathbf{P_{L}}$ the same for all lines; [ACLineX]: $\Delta \mathbf{P_{L}}$ proportional to line reactance.} \label{ctrl_fig:series_plac_3line}
\end{figure}

For the first HVDC placement, the algorithm minimizes $\lVert P_{DC} \rVert_1$. This is heavily influenced by $\lVert CV \rVert_1$ which is presented in Fig.~\ref{ctrl_fig:1normCV_vs_vlmcvxhull_1}. Same to the algorithm presented in Section~\ref{sec:vectorbasedalg}, this algorithm also results in the selection of $CV_{48}$ as the best placement for the first HVDC line.

Figure~\ref{ctrl_fig:series_plac_2line} presents the results for the second HVDC placement based on definition \eqref{ctrl_equ:def_controllability_alternative}. We observe here that all three approaches for setting $\Delta \mathbf{P_{L}}$ result in similar ranking for the top five possible locations. Again, both algorithms result in the same placement for the second HVDC line $CV_{17}$. Fig.~\ref{ctrl_fig:series_plac_3line} presents the results for the third HVDC placement. As we see here, the three different $\Delta \mathbf{P_{L}}$ approaches result in similar ranking for the first nine possible locations. On the other hand, we observe that the two placement algorithms, based on alternative definitions, result in different optimal placement locations for the third line.
\vspace{-0.4cm]}
\section{Discussion}
\label{ctrl_sec:plac_algorithm_discussion}
Comparing the two placement algorithms, although their objective functions are different, they can arrive to similar results if the optimal placement fulfills both definitions for controllability. The advantages of the vector-based algorithm is that it requires substantially less computation time and that we do not need to determine different metrics for setting $\Delta\mathbf{P_L}$. The LP-problem, in comparison, requires 4'095 iterations for the second placement and 16'380 iterations for the third placement. On the other hand, the vector-based algorithm requires a proper calibration of the heuristics used (e.g., $cos\phi$ limit). To eliminate redispatching costs with a limited number of controllers, besides controllability, loading patterns and locations and costs of the generators need to be considered. This paper focused on the introduction of the concept and the placement algorithms. Future work will extend these algorithms, and will also investigate how different $\Delta P_{DC}$ setpoints for different lines affect the algorithm performance.

\paragraph*{Placing HVDC lines in parallel}
In Fig.~\ref{ctrl_fig:series_plac_2line}~and~\ref{ctrl_fig:series_plac_3line}, we observe that the magnitude of control actions is zero when placing an additional HVDC line in parallel with an existing one. Such a placement results to an infeasible solution. This is because an arbitrary number of parallel HVDC lines cannot control the flow of more than one AC line in the system. With $\Delta \mathbf{P_L} = \mathbf{CV_{mn}} (P^{DC}_{mn(1)} + P^{DC}_{mn(2)})$ and $P^{DC}_{mn}, P^{DC}_{mn(2)}$ being scalars, by placing two HVDC lines in parallel we maintain only one degree of freedom.

\paragraph*{Application for TCSCs and PSTs}
For TCSCs, instead of the controllability vectors the relationship in \eqref{ctrl_equ:TCSC_thetaPl} can be used, where $\mathbf{CV_{ij}^{\textnormal{TCSC}}} = \frac{\partial \mathbf{P_L} }{\partial x_{ij}}$ and $\Delta\mathbf{P_L} = \mathbf{CV_{ij}^{\textnormal{TCSC}}} \cdot \Delta x_{ij}$. The difference is that in the case of TCSCs, \eqref{ctrl_equ:TCSC_thetaPl} is dependent on the operating point -- the vector $\mathbf{P_B}$ is part of the equation. As a result, the placement algorithm might need to take several instances of \eqref{ctrl_equ:TCSC_thetaPl} into account for each pair of nodes, considering different operating points. For PSTs, an expression similar to \eqref{ctrl_equ:TCSC_thetaPl} can be derived, where the derivative to be computed will be with respect to the angle change instead of the reactance change. Again, it is expected that this vector will depend on the operating point as with TCSCs.
\vspace{-0.2cm}
\section{Conclusions and Outlook}
In this paper we introduced the concept ``A Fully Controllable Power System'', which aims to fully decouple the market operations from the security considerations in power systems. In a fully controllable system, the electricity market could freely determine the generator dispatch, keeping the redispatching actions which maintain system security to a minimum. In comparison with current AC power systems, controllable line flows eliminate the dependency on the injection points. Thus, in case of contingency the power can be rerouted so that all load is served and no overloadings occur. This minimizes the redispatching costs, and in effect eliminates the Cost of Security.

We have dealt with three main topics in this paper. First, we determined the upper bound on the number of controllable lines in any system, which is $n_L-n_B+1$ for placing controllers in series and $n_B-1$ for placing controllers in parallel with existing AC lines ($n_L$ is the number of lines and $n_B$ is the number of nodes). Showing that each controller can control a single AC line, the same expressions serve as lower bounds for the necessary number of controllers to make a power system ``fully controllable''. Second, we intoduced the controllability vector, as a measure of the controller effect on any AC line flow. Third, we used this vector to formulate two controller placement algorithms that maximize controllability and compare their results. The formulations depend on two alternative definitions of controllability.

Future work includes the formulation and demonstration of a convex optimization algorithm based on the controllability vector, alternative to the algorithm presented in~\ref{sec:vectorbasedalg}. 

\vspace{-0.4cm}
\section{Acknowledgements}
The authors would like to thank Prof. Damien Ernst and Prof. Seth Blumsack for the helpful discussions.
\vspace{-0.4cm}
\bibliographystyle{IEEEtran}
\bibliography{fully_controllable_system_biblio}

%
%
%




\end{document}